\documentstyle[12pt]{article}

\newcommand{\del}{\partial}

\newcommand{\eqn}[1]{(\ref{#1})}
\newcommand{\complex}{{\bb C}} 
\newcommand{\zed}{{\bb Z}} 
\newcommand{\real}{{\bb R}} 
\newcommand{\id}{{\bb I}} 
\newcommand{\alg}{{\cal A}} 

\font\mybb=msbm10 at 12pt
\def\bb#1{\hbox{\mybb#1}}

\def\e{{\rm e}}
\def\slash{\!\!\!\!/}
\def\Dirac{{D\!\!\!\!/\,}}
\def\beq{\begin{equation}}
\def\eeq{\end{equation}}
\def\bea{\begin{eqnarray}}
\def\eea{\end{eqnarray}}

\def\bd{\begin{displaymath}}
\def\ed{\end{displaymath}}

\makeatletter
\newdimen\normalarrayskip              
\newdimen\minarrayskip                 
\normalarrayskip\baselineskip
\minarrayskip\jot
\newif\ifold             \oldtrue            \def\new{\oldfalse}
\def\arraymode{\ifold\relax\else\displaystyle\fi} 
\def\@arrayskip{\ifold\baselineskip\z@\lineskip\z@
     \else
     \baselineskip\minarrayskip\lineskip2\minarrayskip\fi}
\def\@arrayclassz{\ifcase \@lastchclass \@acolampacol \or
\@ampacol \or \or \or \@addamp \or
   \@acolampacol \or \@firstampfalse \@acol \fi
\edef\@preamble{\@preamble
  \ifcase \@chnum
     \hfil$\relax\arraymode\@sharp$\hfil
     \or $\relax\arraymode\@sharp$\hfil
     \or \hfil$\relax\arraymode\@sharp$\fi}}
\def\@array[#1]#2{\setbox\@arstrutbox=\hbox{\vrule
     height\arraystretch \ht\strutbox
     depth\arraystretch \dp\strutbox
     width\z@}\@mkpream{#2}\edef\@preamble{\halign \noexpand\@halignto
\bgroup \tabskip\z@ \@arstrut \@preamble \tabskip\z@ \cr}%
\let\@startpbox\@@startpbox \let\@endpbox\@@endpbox
  \if #1t\vtop \else \if#1b\vbox \else \vcenter \fi\fi
  \bgroup \let\par\relax
  \let\@sharp##\let\protect\relax
  \@arrayskip\@preamble}
\makeatother

\newlength{\extraspace}
\newlength{\extraspaces}
\begin{document}

\renewcommand{\footnotesize}{\small}

\addtolength{\baselineskip}{.8mm}

\thispagestyle{empty}

\begin{flushright}
\baselineskip=12pt
OUTP-97-47P\\
hep-th/9709198\\
\hfill{  }\\ September 1997
\end{flushright}
\vspace{.5cm}

\begin{center}
\baselineskip=12pt

{\large\sc{Electric-magnetic Duality in Noncommutative Geometry}}\\[20mm]

{\sc Fedele Lizzi}$^{a,b,}$\footnote{Email: {\tt lizzi@na.infn.it}} {\sc
and Richard J.\ Szabo}$^{b,}$\footnote{Work supported in part by the Natural
Sciences and Engineering Research Council of Canada.\\ Email: {\tt
r.szabo1@physics.oxford.ac.uk}} \\[6mm]

$^a$ {\it Dipartimento di Scienze Fisiche, Universit\`a di Napoli Federico II\\
and INFN, Sezione di Napoli, Italy} \\[3mm]

$^b$ {\it Department of Physics -- Theoretical Physics, University of Oxford\\
1
Keble Road, Oxford OX1 3NP, U.K.} \\[15mm]

\vskip 1.0 in

{\sc Abstract}

\begin{center}
\begin{minipage}{14cm}

The structure of S-duality in $U(1)$ gauge theory on a 4-manifold $M$ is
examined using the formalism of noncommutative geometry. A noncommutative space
is constructed from the algebra of Wilson-'t Hooft line operators which
encodes both the ordinary geometry of $M$ and its infinite-dimensional loop
space geometry. S-duality is shown to act as an inner automorphism of the
algebra and arises as a consequence of the existence of two independent Dirac
operators associated with the spaces of self-dual and anti-selfdual 2-forms on
$M$. The relations with the noncommutative geometry of string theory and
T-duality are also discussed.

\end{minipage}
\end{center}

\end{center}

\vfill
\newpage
\pagestyle{plain}
\setcounter{page}{1}

One of the oldest forms of duality in physics is electric-magnetic duality
which asserts the invariance of the vacuum Maxwell equations under the
interchange of electric and magnetic fields. This phenomenon is an example of
an `S-duality' which originally appeared in the context of four-dimensional
abelian gauge theories \cite{cardy}, such as electrodynamics with instanton
term\footnote{Here and in the following we shall not write explicit metric
factors and all quantities are implicitly covariant.}
\beq
S[A]=\frac1{16\pi}\int_Md^4x~\left(-\frac{4\pi}{g^2}F_{\mu\nu}F^{\mu\nu}+
\frac{i\theta}{2\pi}\epsilon^{\mu\nu\lambda\rho}F_{\mu\nu}
F_{\lambda\rho}\right)
\label{maxcomp}\eeq
defined on a four-dimensional Euclidean spacetime manifold $M$, where $g$ is
the electromagnetic coupling constant and $\theta$ is the vacuum angle.
S-duality in this case implies the invariance of the quantum field theory under
the modular transformation $\tau\to-1/\tau$ of the parameter
\beq
\tau=\frac\theta{2\pi}+\frac{4\pi i}{g^2}
\label{taudef}\eeq
which lives in the upper complex half-plane. It therefore relates the strong
and weak coupling regimes of the quantum field theory. This duality extends to
$N=4$
supersymmetric Yang-Mills theory \cite{vw}, and recently this property has been
exploited to obtain non-perturbative solutions of $N=2$ supersymmetric quantum
chromodynamics \cite{sw} leading to a picture of quark confinement in terms of
the condensation of magnetic monopoles. S-duality has also enabled much
progress towards the exact solution of $N=1$ supersymmetric gauge theories
\cite{seiberg}.

Many generic features of S-duality can be learned from the simple
quantum field theory \eqn{maxcomp}. The global duality properties of this model
have been studied from a path integral point of view in \cite{global} and in
terms of canonical transformations on the gauge theory phase space in
\cite{lozano}. In this letter we will show how the duality properties of the
theory \eqn{maxcomp} naturally emerge within the formalism of noncommutative
geometry \cite{connes}. We will show that the same features of the
noncommutative spacetime formulation of target space duality in string theory
\cite{ls} emerge in this description. This extends the worldsheet formulations
of the noncommutative geometry of string spacetimes \cite{ls}--\cite{fgr} to a
genuine spacetime description. These constructions could have strong
implications on the unified structure of M Theory \cite{M} which relates
different spacetime theories to one another via target space and S-duality
transformations. The low-energy dynamics of M Theory has been recently
conjectured to be described by a dimensionally-reduced supersymmetric
Yang-Mills theory \cite{bfss}. A noncommutative geometry formalism for this
description has been discussed in \cite{Mncg} (see also \cite{fgr}).

The basic object which describes a metric space in noncommutative geometry is
the spectral triple $({\cal A},{\cal H},D)$, where ${\cal A}$ is a $*$-algebra
of bounded operators acting on a separable Hilbert space $\cal H$ and $D$ is a
(generalized) Dirac operator on $\cal H$. A spin-manifold $M$ with metric
$g_{\mu\nu}$ is described by the choice of ${\cal H}=L^2(M,S)$, the space of
square-integrable spinors on $M$ (i.e. $L^2$-sections of the spin bundle
$S(M)$), and the {\it abelian} algebra $\alg=C^\infty(M,\complex)$ of smooth
complex-valued functions on $M$ acting by pointwise multiplication in $\cal H$.
This is the canonical $*$-algebra associated with any manifold, and it
determines the topology and differentiable structure of a space through the
smoothness criterion. In fact, there is a one-to-one correspondence between the
set of all Hausdorff topological spaces and the collection of commutative
$C^*$-algebras, and therefore the study of the properties of spacetime
manifolds can be substituted by a study of the properties of abelian
$*$-algebras. The usual Dirac operator $D=ig^{\mu\nu}\gamma_\mu\nabla_\nu$ then
describes the
Riemannian geometry of the manifold, where the real-valued gamma-matrices obey
the Clifford algebra $\{\gamma_\mu,\gamma_\nu\}=2g_{\mu\nu}$ and $\nabla_\mu$
is the covariant derivative constructed from the spin-connection. Thus,
roughly speaking, $D$ is the ``inverse" of the infinitesimal $dx$ which
determines geodesic distances in the spacetime. It encodes the Riemannian
geometry via both the gamma-matrices and the spin-connection. Note that this
spectral triple naturally arises from quantizing the free geodesic motion of a
test particle on $M$. In that case $\cal H$ is the Hilbert space of physical
states, $\alg$ is the algebra of observables, and the Hamiltonian $H=-D^2$ is
the Laplace-Beltrami operator. One power of this description is the possible
generalization to {\it noncommutative} algebras $\alg$ leading to the notion of
a ``noncommutative space", such as that anticipated from string theory
\cite{ls}.

In the following we will construct a spectral triple appropriate for the
description of the duality properties of the abelian gauge theory
\eqn{maxcomp}. The ``space" is described by the noncommutative algebra of
Wilson-'t Hooft line operators which incorporates the ordinary (commutative)
spacetime geometry of the 4-manifold $M$, its infinite-dimensional loop space
geometry, and also the geometry of the gauge theory \eqn{maxcomp}. We describe
the similarities between this algebra and the vertex operator algebras that are
used in the construction of string spacetimes \cite{ls}--\cite{fgr} and also
those of M Theory \cite{fgr,Mncg}. We shall show that the duality symmetries
arise in the same way that target space duality did in \cite{ls}. Namely, the
``chiral" structure of the theory, determined by the decomposition of the space
of 2-forms on $M$ into self-dual and anti-selfdual parts, implies the existence
of two independent Dirac operators for the noncommutative geometry. Duality is
then represented as a change of Dirac operator which is just a change of metric
for the noncommutative geometry, i.e. an isometry of the ``space". Furthermore,
S-duality is represented as an inner automorphism of the $*$-algebra $\alg$,
meaning that in this context it is a gauge transformation representing internal
fluctuations of the geometry. We also discuss the similarities between the
noncommutative geometries for gauge fields and strings, illustrating the
unified description of duality and the spacetime structure of string theory
that emerges from the formalism of noncommutative geometry.

Before constructing the appropriate spectral triple, we shall need some basic
results concerning the geometry of 4-manifolds. Consider the $U(1)$ gauge
theory \eqn{maxcomp} on the 4-manifold $M$, which we assume is closed and
admits a spin structure (equivalently it has a vanishing second Stiefel-Whitney
class). We choose a complex line bundle $L\to M$ and a gauge connection 1-form
$A$ on $L$ with curvature $F$. The curvature obeys the Bianchi identity $dF=0$
and has integer period around 2-cycles $\Sigma\subset M$, $\oint_\Sigma
F/2\pi\in{\bb Z}$. Picking a conformal equivalence class of Euclidean signature
metrics $g_{\mu\nu}$ on $M$, we can define a Hodge star-operator $\star$ on $M$
and consider the 2-form $\star F$ dual to the curvature $F$. From this we
define the self-dual and anti-selfdual decompositions
\beq
F^\pm=\mbox{$\frac12$}(F\pm\star F)
\label{dualdecomp}\eeq
according to the splitting of the vector space of 2-forms on $M$ as
$\Omega^2(M)=\Omega^2_+(M)\oplus\Omega_-^2(M)$. The line bundle $L$ is
classified topologically by the first Chern class
\beq
p^\perp\equiv c_1(L)=[F/2\pi]\in H^2(M;{\bb Z})
\label{chernclass}\eeq
For classical field configurations $p^\perp$ is a harmonic form,
$dp^\perp=d\star p^\perp=0$, and the closed 2-form $F$ can be written using the
Hodge decomposition as
\beq
F=dA+2\pi p^\perp
\label{hodgeF}\eeq
where $A$ is a single-valued 1-form. For a given complex line bundle $L$, the
connections $A$ obeying \eqn{hodgeF} live in the torus
$H^1(M;\real)/H^1(M;\zed)$, where the real cohomology classes take into account
all canonical locally gauge-equivalent connections, while the integer
cohomology accounts for the equivalence under large gauge transformations.

The second cohomology group $H^2(M;{\bb Z})$ carries the intersection form
\beq
G(\alpha,\beta)=\int_M\alpha\wedge\beta~~~~~~\mbox{for}~~\alpha,\beta\in
H^2(M;{\bb Z})
\label{intform}\eeq
Any closed surface $\Sigma\subset M$ can be expressed in terms of a canonical
basis $\Sigma^a$ of homology 2-cycles. The inverse $G^{ab}$ of the intersection
matrix $G_{ab}\equiv G(\alpha_a,\alpha_b)$, with $\alpha_a$ a basis of
$H^2(M;{\bb Z})$ dual to $\Sigma^a$, i.e.
$\oint_{\Sigma^a}\alpha_b=\delta_{ab}$, is the signed intersection number
$G^{ab}=\Sigma^a\cap\Sigma^b$ of the homology 2-cycles. The intersection form
(\ref{intform}) of the 4-manifold $M$ is always symmetric and non-degenerate,
so that the matrix $G_{ab}$ determines a quadratic form on the lattice
$H^2(M;\zed)$. Since $M$ is a spin manifold, $H^2(M;{\bb Z})$ is even, and by
Poincar\'e duality $H^2(M;{\bb Z})$ is self-dual. Thus the second cohomology
group of the spacetime equipped with the inner product $G_{ab}$ defines an even
self-dual Lorentzian lattice $\Lambda=H^2(M;{\bb Z})=H_+^2(M;{\bb Z})\oplus
H_-^2(M;{\bb Z})$. Its rank is the second Betti number $b_2=\dim H^2(M)$ and
its signature is $(b_2^+,b_2^-)$, with $b_2^\pm=\dim H^2_\pm(M)$ the dimensions
of the spaces of self-dual and anti-selfdual harmonic 2-forms, respectively.

The Maxwell action (\ref{maxcomp}) with instanton term can be written as
\beq
S[A]=\frac i{4\pi}\int_M\left(\bar\tau F^+\wedge F^++\tau F^-\wedge
F^-\right)=\frac i{4\pi}\int_MF\wedge\tilde F
\label{maxwell}\eeq
where
\beq
\tilde F=\frac\theta{2\pi}F+\frac{4\pi i}{g^2}\star F=\tau F^++\bar\tau F^-
\label{dualstrength}\eeq
is the dual field strength. It is evident from \eqn{maxwell} that
$F$ and $\tilde F$, or alternatively $F^+$ and $F^-$, play a symmetric role in
the theory. The classical equations of motion are
$d\star F=0$, which combined with the Bianchi identity yield the Maxwell
equations of electrodynamics. This implies that $d\tilde F=0$, so that by
Poincar\'e's lemma, there exists locally a one-form $\tilde A$ on $M$ which is
a potential for the dual field strength, $\tilde F=d\tilde A+\tilde p^\perp$
(if $M$ is a simply-connected 4-manifold then this is true globally). The dual
gauge field $\tilde A$ is related to $A$ by
\beq
\tilde A=\frac{4\pi i}{g^2}d^{-1}\star dA+\frac\theta{2\pi}A+k^\perp
\label{dualA}\eeq
where $k^\perp$ is a harmonic one-form. But other than this formal non-local
expression, there is no simple relationship between $A$ and $\tilde A$.
Moreover, their simultaneous existence ceases in the presence of magnetic
or electric sources. Nevertheless, when both electric and magnetic charges are
present (or both absent), $A$ and $\tilde A$ are equivalent, and one could use
either one as a configuration space variable. Here we will consider both of
them on equal footing as independent variables in an enlarged algebra. We
will then project onto a space with only one potential by imposing the physical
constraints represented by Gauss' law.

To study the quantum field theory, we employ canonical quantization. For this,
we work in a three-dimensional spatial slice of the 4-manifold $M$. Locally, in
this region, the spacetime is the product manifold $M=\real\times M_3$, where
$M_3$ is a compact 3-manifold without boundary, and in this region we fix the
temporal gauge $A_0=\tilde A_0=0$. Corresponding to $A$ and $\tilde A$ there
are two canonically conjugate momenta
\beq
\Pi_i={\delta S\over\delta\del_0A^i}=-4\star\tilde F_{0i}=4\left(\bar\tau
F_{0i}^+-\tau F_{0i}^-\right)~~~,~~~\tilde \Pi_i={\delta
S\over\delta\del_0\tilde A^i}=4\star F_{0i}=4\left(F_{0i}^+-F_{0i}^-\right)
\eeq
with the second class constraints $\Pi_0\sim0,\tilde\Pi_0\sim0$, and we define
$\Pi_i^\pm=2(\tau-\bar\tau)F_{0i}^\pm$. The generator of local gauge
transformations is given by the variation\footnote{Note that we could instead
use the generator $\tilde{\cal G}$ in the following defined by varying the dual
field component $\tilde A_0$. This invariance under $A\leftrightarrow\tilde A$
applies to all of our constructions which follow, and it is the essence of the
duality properties which we shall exhibit here.}
\beq
{\cal G}=\frac{\delta S}{\delta A_0}=\frac i\pi\partial^i\left(\bar\tau
F_{0i}^+-\tau F_{0i}^-\right)=\frac i{4\pi}\partial^i\Pi_i
\label{gaugegen}\eeq
and Gauss' law is the second class constraint ${\cal G}\sim0$ to be imposed on
the physical state space of the gauge theory. The Hamiltonian can be written as
\beq
H=\int_{M_3}d^3x~\frac1{4(\tau-\bar\tau)}\Pi^{+i}\Pi_i^-
\label{Ham}\eeq

To construct the Hilbert space of the gauge theory, we quantize the classical
field configurations. For this, we adopt a functional Schr\"odinger
polarization on the extended gauge theory configuration space
$(A_+,A_-)\equiv(A-\frac1{\bar\tau}\tilde A,A-\frac1\tau\tilde A)$ with the
conjugate momenta $\Pi_i^\pm=-i\frac\delta{\delta A_\pm^i}$, so that the
physical states are the wavefunctionals $\Psi[A_+,A_-]\equiv|A_+,A_-\rangle$.
The oscillator form of the Hamiltonian \eqn{Ham} suggests to employ a coherent
state quantization and take the wavefunctionals to be eigenstates of the
momentum operators
\beq
\Pi_i^+(x)|A_+,A_-\rangle=\bar\tau\Lambda_i^+(x)|A_+,A_-\rangle~~~~,~~~~
\Pi_i^-(x)|A_+,A_-\rangle=\tau\Lambda_i^-(x)|A_+,A_-\rangle
\label{cohcondn}\eeq
where $\Lambda^\pm$ are one-forms on $M_3$. These equations are solved by
\beq
|A_+,A_-\rangle=\exp\left[i\int_{M_3}d^3x~\left(\bar\tau\Lambda_i^+(x)
A_+^i(x)+\tau\Lambda_i^-(x)A_-^i(x)\right)\right]
\label{Psisoln}\eeq
The Gauss' law constraint is
\beq
{\cal G}(x)|A_+,A_-\rangle=\frac1{4\pi(\bar\tau-\tau)}\partial^i\left(\tau\frac
\delta{\delta A_+^i(x)}-\bar\tau\frac\delta{\delta
A_-^i(x)}\right)|A_+,A_-\rangle=0
\label{gausscondn}\eeq
which implies that $\Lambda_i^+(x)=\Lambda_i^-(x)+k_i^\perp(x)$, where
$k^\perp$ is a harmonic one-form on $M_3$, $\partial_ik_i^\perp=0$,
representing the same degree of freedom as in \eqn{dualA}. Thus the
wavefunctionals, with momentum and harmonic quantum numbers $\Lambda$ and
$k^\perp$, are given by
\beq
|A_+,A_-;\Lambda,k^\perp\rangle=\exp\left[i\int_{M_3}d^3x~\Lambda_i(x)
\left(\bar\tau A_+^i(x)+\tau
A_-^i(x)\right)+i\int_{M_3}d^3x~k_i^\perp(x)A_+^i(x)\right]
\label{wavesolns}\eeq
The wavefunctionals \eqn{wavesolns} are eigenstates of the Hamiltonian
\eqn{Ham} with the energy eigenvalues
\beq
{\cal E}_{\Lambda,k^\perp}=\int_{M_3}d^3x~\frac\tau
{4(\tau-\bar\tau)}\Lambda^i(x)\left(\bar\tau\Lambda_i(x)+k_i^\perp(x)\right)
\label{spectrum}\eeq

The Hilbert space ${\cal H}_A$ spanned by the states \eqn{wavesolns} can be
used to represent a unital $*$-algebra of observables appropriate to the gauge
theory. The basic observables describing ``interactions" in the theory are the
gauge- and topologically-invariant Wilson line operators. In terms of the
electric and magnetic charges $q_e$ and $q_m$, the fundamental holonomy
operators are
\beq
W_{q_e}^{(C)}[A]=\exp\left(iq_e\oint_CA\right)~~~~~~,~~~~~~\tilde
W_{q_m}^{(C)}[\tilde A]=\exp\left(iq_m\oint_C\tilde A\right)
\label{wilsonem}\eeq
where $C\subset M$ are loops. Single-valuedness of these operators under large
gauge transformations yields the Dirac quantization condition $q_e,q_m\in\zed$.
The action of the Wilson loop operators on the Hilbert space can be computed by
combining the two operators \eqn{wilsonem}, using the Baker-Campbell-Hausdorff
formula, into the abelian Wilson-'t Hooft line operators
\beq
W_{q_e,q_m}^{(C)}[A,\tilde A]=\e^{-i\pi q_eq_mL(C)}~W_{q_e}^{(C)}[A]\tilde
W_{q_m}^{(C)}[\tilde A]=\exp\left[i\oint_C\left(q_+A_++q_-A_-\right)\right]
\label{wilsoncomb}\eeq
where $L(C)$ is the self-linking number of the loop $C$ and $q_+=q_e+\tau q_m$,
$q_-=q_e+\bar\tau q_m$. They describe the interaction of electric charges and
magnetic monopoles with the electromagnetic field. We can write the contour
integrals in \eqn{wilsoncomb} as
\beq
\oint_CA_\pm=\int_Md^4x~A_\pm^\mu(x)J_\mu^{(C)\perp}(x)
\label{contcurrent}\eeq
where $J_\mu^{(C)\perp}(x)=\int_0^1ds~\dot
x^{(C)}_\mu(s)\delta^{(4)}(x-x^{(C)}(s))$ is the conserved current of the
closed worldline $C$, with $s\in[0,1]$ the parameter of $C$ and
$x^{(C)}(s):[0,1]\to M$ the embedding of $C$ in $M$. From this it follows that
the action of the operators \eqn{wilsoncomb} on the wavefunctionals
\eqn{wavesolns} yields
\beq
W_{q_e,q_m}^{(C)}[A,\tilde A]|A_+,A_-;\Lambda,k^\perp\rangle=
|A_+,A_-;\Lambda(C),k^\perp(C)\rangle
\label{wilsonHaction}\eeq
where $\Lambda(C)=\Lambda+\frac1{\bar\tau}q_-J^{(C)\perp}$ and
$k^\perp(C)=k^\perp+[(1+\frac1\tau)q_+-\frac1{\bar\tau}q_-]J^{(C)\perp}$.

The algebra of the Wilson-'t Hooft operators \eqn{wilsoncomb} can also be
computed. Assuming that the loop $C$ is contractible (this is immediate if $M$
is simply-connected) we can use Stokes' theorem to rewrite the line integrals
of $A_\pm$ as surface integrals of the curvatures $F^\pm$ over a surface
$\Sigma(C)$ spanned by $C$, i.e. $\partial\Sigma(C)=C$. A simple calculation
shows that these operators then obey the clock algebra
\beq
W_{q_e,q_m}^{(C)}[A,\tilde A]W_{q_e',q_m'}^{(C')}[A,\tilde
A]=\e^{q_+L(C,C')q_-'/2}~W_{q_e',q_m'}^{(C')}[A,\tilde A]
W_{q_e,q_m}^{(C)}[A,\tilde A]
\label{wilsonclock}\eeq
where
\beq
L(C,C')=\int_{\Sigma(C)}\int_{\Sigma(C')}\left[F^+,F^-\right]
\label{intnum}\eeq
defines a local intersection number of the curves $C$ and $C'$. In fact, the
representation of the operators \eqn{wilsoncomb} in terms of surfaces
$\Sigma\subset M$, i.e.
$W_{q^+,q^-}(\Sigma)=\exp[i\int_\Sigma(q^+F^++q^-F^-)]$, emphasizes the fact
that they are actually defined in terms of the elements $(F^+,F^-)$ of the
even, self-dual Lorentzian lattice $\Lambda=H^2(M;\zed)$. For closed surfaces
$\Sigma$, these operators depend only on the harmonic components $p_\pm^\perp$
of the field strengths, and their clock algebra can be expressed in terms of
the intersection matrix as
\beq
W_{q_a^+,q_a^-}(\Sigma^a)W_{q_b^+,q_b^-}(\Sigma^b)=\e^{2\pi
iq_a^+G^{ab}q_b^-}~W_{q_b^+,q_b^-}(\Sigma^b)W_{q_a^+,q_a^-}(\Sigma^a)
\label{wilsonsurfalg}\eeq

The Wilson-'t Hooft operators \eqn{wilsoncomb} form a basis for a
noncommutative, infinite-dimensional unital $*$-algebra ${\cal A}_A$. It
describes topological and geometrical properties of the 4-manifold $M$, as well
as the geometry of the complex line bundle $L\to M$. The construction of an
algebra $\alg_A$ from an even self-dual lattice $\Lambda$ was a crucial
ingredient in the description of noncommutative string spacetimes in \cite{ls}.
In that case, the spacetime was described by the vertex operator algebra of the
underlying conformal field theory. Here we see that the algebra $\alg_A$ has a
similar structure, in terms of both its definition and its algebraic
relations. This sets a unified framework for the descriptions of both gauge
theories and string theories using noncommutative geometry.

The algebra \eqn{wilsonsurfalg} is actually a large generalization of one of
the original examples of a noncommutative geometry, the noncommutative torus
\cite{connes,Rieffel}. This algebra describes the quotient of the torus by the
orbit of a free particle whose velocity forms an irrational angle with respect
to the cycles of the torus. In this case the motion is ergodic and dense in the
torus, and the resulting quotient is not a topological space in the usual
sense. An equivalent way to visualize this is to consider the irrational
rotations of a circle. It is then possible to describe the space by the algebra
of functions on a circle together with the action of these irrational
rotations. Such an algebra is generated by two elements $U$ and $V$ which obey
\beq
UV=\e^{i\pi\alpha}~VU
\label{nctorus}\eeq
with $\alpha$ the angle of rotation. This algebra has also appeared in the
recent matrix model descriptions of M Theory \cite{fgr,bfss}. It is possible to
prove \cite{Rieffel} that the algebra generated by $\alpha$ and the ones
generated by $\alpha+1$ and $1/\alpha$ are equivalent. The algebra
\eqn{wilsonsurfalg} is much larger than that of \eqn{nctorus}, and the role of
the irrationality of $\alpha$ translates into a similar condition on $\tau$. It
is, however, very suggestive to notice that in our construction, for each
choice of the charges $q^\pm$, the elements of the second homology group form a
(larger dimensional) noncommutative torus.

The algebra $\alg_A$ is also intimately related to the loop space geometry of
the manifold $M$. A generic algebra element is a linear combination given by
the quantum mechanical path integral
\beq
W_{q_e,q_m}[A,\tilde A]=\int_{C^\infty(M,S^1)}[dx^{(C)}(s)]~\widehat{\cal
W}[x^{(C)}(s)]~W_{q_e,q_m}^{(C)}[A,\tilde A]
\label{genalgelt}\eeq
over the loop space $C^\infty(M,S^1)$ of $M$, where $\widehat{\cal W}$ is a
functional of only the worldline embeddings $x^{(C)}(s)$. Symbolically, we then
have the diagram
\beq\new{\begin{array}{ccc}
C^\infty(M,S^1)&{\buildrel
W^{(\cdot)}\over\longrightarrow}&\alg_A\\{\scriptstyle\pi}\downarrow&
&\downarrow{\scriptstyle\pi_{\alg_A}}\\M&{\buildrel\chi_\cdot\over
\longrightarrow} &C^\infty(M,\complex)\end{array}}
\label{loopspdiag}\eeq
In this diagram the Wilson loops are regarded as functions on the loop space,
which is viewed as an infinite-dimensional vector bundle over the manifold $M$
with bundle projection $\pi$, where $\pi^{-1}(x)$ is the space of loops based
at $x\in M$. The projection $\pi_{\alg_A}$ is defined by restricting the Wilson
loops to constant paths (i.e. loops of minimal area zero), which in turn
projects the algebra $\alg_A$ onto the commutative algebra
$C^\infty(M,\complex)$ describing the ordinary spacetime geometry of the
4-manifold $M$. We shall describe this projection somewhat more precisely
below. For each $x\in M$, there corresponds the character $\chi_x(f)=f(x)$ of
the algebra $C^\infty(M,\complex)$ which ``reconstructs" the points, the
topology, and the differentiable structure of the manifold $M$ in purely
algebraic terms \cite{connes}. It is in this sense that the noncommutative
algebra $\alg_A$ is related to the usual geometry of four-dimensional
electrodynamics and the manifold $M$, and moreover this point of view shows
precisely what sort of geometry the noncommutative space here represents.

As discussed in \cite{fgr}, and as has been employed extensively in the case of
string theory in \cite{ls}--\cite{cham} and for D-brane field theory in
\cite{Mncg}, the Dirac operator for the noncommutative space is obtained from
the supercharges in an appropriate supersymmetrization of the bosonic field
action. When the field theory is a two-dimensional sigma-model, these operators
project onto the Dirac-Ramond operators which describe the DeRham complex of
the given target manifold \cite{witten}. In the case at hand, we consider the
$N=1$
supersymmetric abelian gauge theory with action
\beq
S[A,\psi,\bar\psi]=\frac i{4\pi}\int_M\left(F\wedge\tilde F+{\rm
Im}(\tau)\,\bar\psi\nabla\slash\,\psi\right)
\label{susyaction}\eeq
where $\psi_\alpha$ and $\bar\psi_{\dot\alpha}$, with
$1\leq\alpha,\dot\alpha\leq4$, are fermion fields in the chiral and antichiral
representations of $spin(4)$, respectively. For this model the $N=1$
supersymmetry
\beq\new{\begin{array}{c}
\delta A_\mu=i\bar\varepsilon_{\dot\alpha}\gamma_\mu^{\alpha\dot\alpha}
\psi_\alpha+i\varepsilon_\alpha\gamma_\mu^{\alpha\dot\alpha}\bar
\psi_{\dot\alpha}\\\delta\psi_\alpha=\bar\varepsilon_{\dot\alpha}
\gamma_{\mu\nu}
^{\alpha\dot\alpha}F^{\mu\nu}~~~~,~~~~\delta\bar\psi_{\dot\alpha}=
\varepsilon_\alpha\gamma_{\mu\nu}^{\alpha\dot\alpha}F^{\mu\nu}\end{array}}
\label{N=1susy}\eeq
with $\delta=\varepsilon_\alpha Q^\alpha+\bar\varepsilon_{\dot\alpha}\bar
Q^{\dot\alpha}$, is generated by the action of the supercharges
\beq
Q^\alpha=\gamma_\mu^{\alpha\dot\alpha}\bar\psi_{\dot\alpha}\Pi^\mu~~~~~~,
{}~~~~~~\bar Q^{\dot\alpha}=\gamma_\mu^{\alpha\dot\alpha}\psi_\alpha\Pi^\mu
\label{susycharges}\eeq
on the quantum fields. If we define the generalized Dirac operator
\beq
\Dirac^{\alpha\dot\alpha}=\gamma_\mu^{\alpha\dot\alpha}\otimes\Pi^\mu
\label{Diracop}\eeq
then the supercharges \eqn{susycharges} are generated by its action on the
fermion fields, i.e. $Q^\alpha=\Dirac^{\alpha\dot\alpha}\bar\psi_{\dot\alpha}$
and $\bar Q^{\dot\alpha}=\Dirac^{\alpha\dot\alpha}\psi_\alpha$.

However, there is another Dirac operator that can be introduced, namely that
generated by the dual variables
\beq
\tilde\Dirac=\gamma_\mu\otimes\tilde\Pi^\mu
\label{dualDiracop}\eeq
This choice corresponds to representing the $N=1$ supersymmetry above in terms
of the dual gauge field configurations $\tilde A$. Correspondingly, we also
introduce the self-dual and anti-selfdual Dirac operators
\beq
\Dirac^\pm={\rm
Im}(\tau)^{-1}~\gamma_\mu\otimes\Pi_\pm^\mu
\label{Diracpm}\eeq
These Dirac operators all act on the Hilbert space ${\cal
H}=L^2(M,S)\otimes{\cal H}_A$. Accordingly, the algebra of Wilson-'t Hooft
operators is augmented to $\alg=C^\infty(M,\complex)\otimes\alg_A$. Since
$\Pi_\pm^\mu\sim-i\frac\delta{\delta A_\mu^\pm}$ on the configuration space of
the gauge theory, the operators \eqn{Diracpm} are indeed the appropriate Dirac
operators for the present noncommutative geometry. Furthermore, they are
related to the Hamiltonian operator \eqn{Ham} by
\beq
\id\otimes H=\int_{M_3}d^3x~\mbox{$\frac{{\rm Im}(\tau)}{4i}$}~\Dirac^+\Dirac^-
\label{HamDirac}\eeq
Thus, in the sense of \eqn{HamDirac}, the ``square" of the Dirac operators
\eqn{Diracpm} coincides with the Hamiltonian, which in turn naturally defines
the appropriate Laplace-Beltrami operator for the Riemannian geometry
\cite{fg,cham}.

The existence of two independent Dirac operators as metrics for the
noncommutative space severely restricts the geometry. Namely, there exists a
unitary transformation $S:{\cal H}\to{\cal H}$ which is an automorphism of the
algebra $\alg$, i.e. $S\alg S^{-1}=\alg$, and maps the two Dirac operators
\eqn{Diracop} and \eqn{dualDiracop} into one another as
\beq
S\,\Dirac\,S^{-1}=-\tilde\Dirac~~~~~~,~~~~~~S\,\tilde\Dirac\,S^{-1}=\Dirac
\label{Sdirac}\eeq
or equivalently $S\Dirac^+S^{-1}=\tau\Dirac^+$,
$S\Dirac^-S^{-1}=\bar\tau\Dirac^-$. This order-4 transformation is simply the
S-duality transformation $\tau\to-1/\tau$ which amounts to an interchange of
electric and magnetic degrees of freedom, $F\to\tilde F$ and $\tilde F\to-F$.
It leaves both the action \eqn{maxwell} and the Hamiltonian \eqn{HamDirac}
invariant. Explicitly, it acts trivially on the spinor parts of $\cal H$ and
$\alg$ and on ${\cal H}_A$ and $\alg_A$ it is defined by
\beq
S\,|A_+,A_-;\Lambda,k^\perp\rangle=|A_+,A_-;-\Lambda,\bar\tau
k^\perp\rangle~~~~~~,~~~~~~S\,W_{q_e,q_m}[A,\tilde
A]\,S^{-1}=W_{-q_m,q_e}[A,\tilde A]
\label{StransfHA}\eeq
In fact, the explicit operator $S$ which implements this transformation is a
gauge transformation of the noncommutative geometry, i.e. an inner automorphism
of the algebra $\alg_A$ which acts as conjugation by the unitary element
$S=\e^{i{\cal F}}\in\alg_A$, where\footnote{Strictly speaking, the unitary
operator determined by \eqn{autogen} is only an element of the algebra of
extended gauge-invariant observables determined by objects of the form
\eqn{contcurrent} with arbitrary conserved currents $J^\mu$. The continuity
equations for the currents defined in \eqn{autogen} follow from the Maxwell
equations of motion. Nevertheless, we take the form of the unitary
transformation generated by \eqn{autogen} to mean that the S-duality
transformation is an inner automorphism of $\alg$.}
\beq
{\cal F}=\frac1{16\pi}\int_{M_3}d^3x~\left(A_i\Pi^i-\tilde
A_i\tilde\Pi^i\right)
\label{autogen}\eeq

The Hilbert space $\cal H$, the algebra $\alg$, and the spectrum of the
Hamiltonian $H$ are invariant under the above transformation. In terms of the
noncommutative geometry, this implies that the two spectral triples determined
by the Dirac operators \eqn{Diracop} and \eqn{dualDiracop} are isomorphic,
\beq
(\alg~,~{\cal H}~,~\Dirac)\cong(\alg~,~{\cal H}~,~\tilde\Dirac)
\label{spectripiso}\eeq
As a change in choice of Dirac operator in the spectral triple is merely a
change of metric from the point of view of the noncommutative geometry, the
S-duality transformation is simply an isometry of the noncommutative space. It
can also be viewed as a symmetry between an
exterior derivative operator d and its adjoint ${\rm d}^\dagger=\star{\rm
d}\star$ with respect to an appropriate $\star$-operator \cite{ls}. These
features, as well as the fact that the S-duality transformation is a gauge
symmetry, is the essence of duality in this formalism.

By construction, the basis wavefunctionals \eqn{wavesolns} of the Hilbert space
are eigenstates of the Dirac operators \eqn{Diracop} and \eqn{dualDiracop}. The
noncommutative space is thus constructed in terms of the spectrum of the Dirac
operators, which is a feature of the spectral action principle of
noncommutative geometry \cite{cham,specaction} which relates isospectral
Riemannian geometries (ones with the same Dirac K-cycles $({\cal H},D)$) to one
another. It is especially instructive to examine closely the zero mode
eigenspaces $\ker\Dirac$ and $\ker\tilde\Dirac$. The isomorphism
\eqn{spectripiso} when restricted to these subspaces of ${\cal H}$ yields a
physical interpretation of the duality symmetry. In $\ker\tilde\Dirac$, the
states, before the imposition of Gauss' law, are arbitrary functionals of $A$.
On the other hand, the states of $\ker\Dirac$ are arbitrary functionals of the
dual gauge field $\tilde A$. This is the usual statement of S-duality when
considered as a functional canonical transformation on the phase space of the
gauge theory \cite{lozano}. The pair $(A,\tilde A)$ are in this sense a
canonically conjugate pair of variables, and either one can be used as a
configuration space coordinate depending on whether one works in the position
or momentum representation.

Similar statements also hold for the algebra of Wilson-'t Hooft operators,
where the zero-mode subspaces of $\alg_A$ are the commutants of the Dirac
operators, i.e. the elements $W\in\alg$ with $[\Dirac,W]=0$ or
$[\tilde\Dirac,W]=0$. The mapping between commutants then corresponds to
interchanging Wilson and 't Hooft line operators. This is the usual physical
consequence of S-duality in gauge theory, i.e. it interchanges electric charges
and magnetic monopoles, which has profound consequences for the
non-perturbative structure of the quantum field theory \cite{sw}. Note that if
one first imposes gauge invariance represented by Gauss' law and then projects
$\alg_A$ onto these commutants, then the constraints become
$J_\mu^{(C)\perp}=0$, i.e. the Wilson loops which are defined on the constant
loops $x^\mu(s)=x_0^\mu\in M~~\forall s\in[0,1]$. The algebra $\alg$ thus
projects onto $C^\infty(M,\complex)$, and this defines the projection
$\pi_{\alg_A}$ depicted in \eqn{loopspdiag}.

There is another ``internal" symmetry of the quantum spacetime that is not
represented as a change of Dirac operator but does lead to non-trivial
dynamical effects in the quantum field theory. This is the symmetry under the
shift $\theta\to\theta+2\pi$ of the vacuum angle (or $\tau\to\tau+1$). It
corresponds to a change of the integer cohomology class of the instanton form
$F\wedge F$, and in the present formulation it can be absorbed by the shift
$\tilde\Pi\to\tilde\Pi+\Pi$, $\Pi\to\Pi$ of the momentum operators (and hence a
redefinition of the Dirac operators). The corresponding spectral triples are
each unaffected by these global redefinitions of all quantities. Note that
these
shifts correspond to reparametrizations of the elements $(\tilde
F,F)\in\Lambda$ and the charges $(q_m,q_e)$ which define the algebra $\alg$.
Thus only integer-valued multiples of $2\pi$ are allowed as vacuum angle
shifts, in order that they leave invariant the algebra $\alg_A$ (equivalently
to preserve the Dirac quantization condition). These two sets of modular
transformations, $\tau\to-1/\tau$ and $\tau\to\tau+1$, generate the duality
group $SL(2,\zed)$ of four-dimensional quantum electrodynamics under which
$\tau$ transforms by linear fractional transformations and the vector $(\tilde
F,F)\in\Lambda$ as a doublet. Recall that these were essentially also the
symmetries of the algebra \eqn{nctorus} of the noncommutative torus.

Thus the noncommutative geometry formulation of duality in gauge theories bears
a remarkable resemblence to that of target space duality in string theory, just
as other descriptions of duality seem to suggest \cite{global,lozano}. In
particular, the gauge theory ``spacetime" is described by an algebra $\alg_A$
which is determined by $b_2^+b_2^-$ moduli parameters that parametrize the
shape of the lattice $\Lambda$ and take values in the Narain moduli space of
$\Lambda$~\cite{ls}
\beq
{\cal M}_{\rm qu}=O(b_2^+,b_2^-;\zed)\setminus O(b_2^+)\times
O(b_2^-)/O(b_2^+,b_2^-)
\label{gaugemodsp}\eeq
where the arithmetic group $O(b_2^+,b_2^-;\zed)={\rm Aut}(\Lambda)$ contains
the group ${\rm Diff}^+(M)$ of orientation preserving diffeomorphisms of the
4-manifold $M$. This is a subgroup of the group ${\rm Diff}(M)$ of (outer)
automorphisms of the algebra $C^\infty(M,\complex)$. The unity between the
gauge field and string noncommutative geometries puts S-duality in string
theory on the same footing as T-duality, as is required by M Theory \cite{M}.
It also agrees with the recent equivalences suggested by Matrix Theory
\cite{bfss} which unifies string theories within a gauge theoretical
description. The algebraic similarities between $\alg_A$ and vertex operator
algebras then suggest a representation of string scattering amplitudes in terms
of Wilson line correlators. The unity of gauge fields and strings, and hence of
the supersymmetric Yang-Mills theory that comprises Matrix Theory, appears to
be contained in their relations to the noncommutative torus.

\bigskip

{\bf Acknowledgements:} {\sc f.l.} gratefully thanks the Theoretical Physics
Department of Oxford University, where this work was begun, for hospitality.
{\sc r.j.s.} thanks N. Dorey and I. Kogan for insightful discussions on
S-duality.

\end{document}